\documentclass[prl,floats,twocolumn,aps,superscriptaddress]{revtex4}
\usepackage{epsfig}

\begin{document}

\title{Mixing structures in the Mediterranean Sea from Finite-Size Lyapunov Exponents}
\author{Francesco d'Ovidio,
Vicente Fern\'andez, Emilio Hern\'andez-Garc\'\i a, and Crist\'obal L\'opez}
\affiliation{Instituto Mediterr\'aneo de Estudios Avanzados IMEDEA (CSIC-UIB),
    Campus Universitat de les Illes Balears,
E-07122 Palma de Mallorca, Spain.}

\begin{abstract}

We characterize horizontal mixing and transport structures in the surface
circulation of the Mediterranean Sea, as obtained from a primitive equation
circulation model. We calculate the Finite Size Lyapunov Exponents (FSLEs) of the
velocity data set, which gives a direct measure of the local stirring. By proper
election of the FSLE parameters, we focus on the mesoscale structures, locating a
number of vortices embedded in an intricate network of high-stretching lines.
These lines control transport in the system. At the edge of the vortices, a dense
tangle of line intersections appears, identifying strong mixing. The spatial
distribution of FSLEs, averaged over one year, allows to classify areas in the
Mediterranean basin according by their mixing activity. The space average of
FSLEs on selected geographical regions gives a measure for quantifying and
comparing the mixing seasonal variability.

\end{abstract}
\maketitle

\section*{Introduction}

Horizontal transport and mixing processes are central to the study of the
physical, chemical, and biological dynamics of the ocean. Correct understanding
and precise modelling of them are relevant from a theoretical viewpoint and
crucial for a range of practical issues, such as plankton dynamics or the fate of
pollutant spills. In this regard, the last few years have seen the appearance of
interesting new developments \cite{lagrangian} on the Lagrangian description of
transport and mixing phenomena, many of them coming from the area of
nonlinear dynamics. Such approaches do not aim at predicting individual tracer
trajectories, but at locating spatial structures that are known from dynamical
systems theory to act as templates for the whole flow \cite{Ottino89,Wiggins92}.
Such structures (attractors, saddles, manifolds, ...) have been used since many
years ago for classifying the evolution of trajectories in abstract dynamical
systems. However, when put in a fluid dynamics context, they gain a new and
direct physical meaning, corresponding for instance to avenues and barriers to
transport, vortex boundaries, or lines of strong stretching. In particular,
chaotic motions such as the ones occurring in the turbulent ocean are
characterized by complex intersection of stretching and contracting 
manifolds at the so called
{\sl hyperbolic points}: regions of fluid initially compact in the proximity of
these points become elongated along the stretching directions and then folded,
leading to the typical filamental and convoluted structures that are common in
satellite pictures of water temperature or chlorophyll, as well as in laboratory
experiments. From the point of view of transport, such regions are characterized
by a strong mixing: trajectories of initially close particles are quickly
separated along the stretching directions, and fluid of different origins is
inserted in between.

Until recently, the power of these novel Lagrangian approaches has been mainly
relegated to mathematical systems or simplified workbench models, since the
required detailed knowledge of the velocity field was not readily available in
real geophysical situations. However, in the last decades the situation has
dramatically changed, with a rapidly increasing amount of data available from
Lagrangian drifters \cite{lagrangian}, satellite measurements \cite{halpern00},
and especially from detailed computer models \cite{computers,dietrich97}. This
has paved the way to a growing number of geophysical applications, among which we
mention studies of the Loop Current in the gulf of Mexico \cite{Kuznetsov2002},
or the characterization of dispersion properties in the Adriatic Sea from drifter
experiments \cite{lacorata01}. The methodology is not restricted to ocean
circulation, but it is also being developed in atmospheric dynamics
\cite{Joseph2002}, and mantle convection \cite{Farnetani2003}. Indeed, new
Lagrangian approaches are very appealing for geophysical applications, since they
can be used as a tool for automatically extracting transport structures
underlying raw Eulerian velocity data.

In this Letter we characterize mixing strength at the mesoscale in different
areas of the Mediterranean Sea by means of a Lagrangian technique, the Finite
Size Lyapunov Exponents method. The technique also identifies dynamical objects
that organize the transport, and relevant coherent structures.

\section*{Numerical Data}
\label{sec:data}

We produce velocity data from the DieCAST ocean  model (\textit{Dietrich for Center Air
Sea Technology}), adapted to the Mediterranean basin with a horizontal resolution
of $1/8$ degrees (approximately $10$ km) and $30$ vertical levels (see
\cite{Fernandez2004} for details). The DieCAST model is a z-level primitive
equation model based on the hydrostatic, incompressible, and rigid lid
approximations. The model has been integrated for $20$ years being forced by
yearly repeating monthly climatological atmospheric forcing. Using such
climatological forcing the model reproduces well the general surface circulation
and many of the important features of the observed annual cycle of the
Mediterranean Sea \cite{Fernandez2004}. Due to the adequate horizontal resolution
and the numerical characteristics of the model, basically high order numerics and
low numerical and physical dissipation \cite{dietrich97}, the numerical
simulations reproduce a great number of mesoscale structures, that are required
for the present study.

In this work we analyze the daily model output of velocity field obtained from
the last three years of simulation and corresponding to the second vertical
horizontal layer (in this way we avoid the strong dependence on wind stress of
the first model layer). This layer has a vertical extent of $11.60\ m$, centered
at a depth of $16\ m$. The two-dimensional horizontal velocity field on this
layer is not exactly incompressible, but it is very close to this situation since
typical vertical velocities in the ocean are four orders of magnitude smaller than horizontal
ones ($10^{-5}$ vs $10^{-1}\ m/s$). Thus, points at which fluid particles diverge
horizontally along particular directions receive fluid along other directions, so
that stretching is linked to mixing. On the other hand, within the Finite-Size
Lyapunov exponent (FSLE) method used in this Paper and presented in the next
sections to estimate transport at the mesoscale, fluid particle trajectories need
to be integrated only while they undergo horizontal displacements of the order of
$100\ km$, i.e. during 1-10 days (see Section of Results). Estimating an
effective or average vertical velocity for this time as the spatial average of
the vertical velocity field in horizontal regions of that size, one obtains an
effective vertical velocity of $0.1-0.7\ m/day$. Thus, during the time of
integration, most particles do not leave the horizontal layer considered. In
consequence, restricting the study to horizontal motion on a single model layer
is a good description of the full transport processes for the space and time
scales relevant to mesoscale processes within the FSLE approach.

\section*{Finite Size Lyapunov Exponents (FSLEs)}
\label{sec:fsle}
A common way to quantify the stretching by advection is by means of the standard
Lyapunov exponents. They are defined as the exponential rate of separation,
averaged over infinite time, of fluid parcels initially separated
infinitesimally. In realistic situations (such as the case of the Mediterranean
Sea where boundaries at finite distance strongly influence the circulation) the
infinite-time limit in the definition makes the Lyapunov exponent a quantity of
limited practical use. Recently, the Finite Size Lyapunov Exponent (FSLE) has
been introduced \cite{aurell97,Artale97} in order to study non-asymptotic
dispersion processes, which is particularly appropriate to analyze transport in
closed areas. FSLEs have been used for two complementary goals: for
characterizing dispersion processes \cite{lacorata01}, and for detecting and
visualizing Lagrangian structures (e.g., transport barriers or vortex boundaries)
\cite{koh02}. Here we will focus mainly in the second use, but we will also
introduce measures of dispersion and mixing based on the Lagrangian structures
detected.

The FSLE technique appears to be ideally suited for oceanographic applications,
being the mathematical analogous of a floater experiment: a set of tracers, with
some initial mutual distances, are followed in time as they are transported by
integrating the velocity field (we use a bilinear interpolation to assign
velocities to points that are not model grid points; thus the velocity field is
effectively smooth at scales below $1/8$ of degree). The FSLE is inversely
proportional to the time at which two tracers reach a prescribed separation. More
precisely, $\lambda ({\bf x},t, \delta_0,\delta_f)$, the FSLE at position ${\bf
x}$ and time $t$, is computed from the time $\tau$ it takes for a trajectory
starting at time $t$ at a distance $\delta_0$ from ${\bf x}$ to reach a
separation $\delta_f$ from the reference trajectory that started at ${\bf x}$:

\begin{equation} \lambda ({\bf x},t, \delta_0,\delta_f) \equiv
\frac{1}{\tau}\log\frac{\delta_f}{\delta_0}.
\label{fsle}
\end{equation}

In order to characterize the strongest separation (and the fastest convergence
along the complementary direction), $\lambda$ is selected as the maximum among
the four values obtained when the initial separation $\delta_0$ is chosen along
four orthogonal directions.

The FSLE depends on the choice of two length scales: the initial separation
$\delta_0$ and the final one $\delta_f$. Here, we are interested in the spatial
distribution of FSLEs, and thus we calculate them at points $\bf x$ located on a
grid of spacing $\Delta x$. In this case, a simple argument shows the convenience
of using a value of $\delta_0$ close to the intergrid spacing $\Delta x$:  If one
chooses $\delta_0$ much smaller than $\Delta x$, all the points of a stretching
manifold laying further than $\delta_0$ from any grid point are not tested, and
thus the method gives only a rather discontinuous sampling of the structure. On
the other hand, if $\delta_0$ is much larger than $\Delta x$, the same stretching
manifold is detected (``smeared'') on several grid points, with a loss in spatial
resolution. Since we are interested in mesoscale structures, the other length,
$\delta_f$, will be chosen as $\delta_f=1$ degree, i.e., separations of about
$110$ Km. In this way the FSLE represents the inverse time scale for mixing up
fluid parcels between length scales  $\delta_0$ and  $\delta_f$.

\section*{Results}
\label{sec:results}

\begin{figure}
\includegraphics[width=39pc]{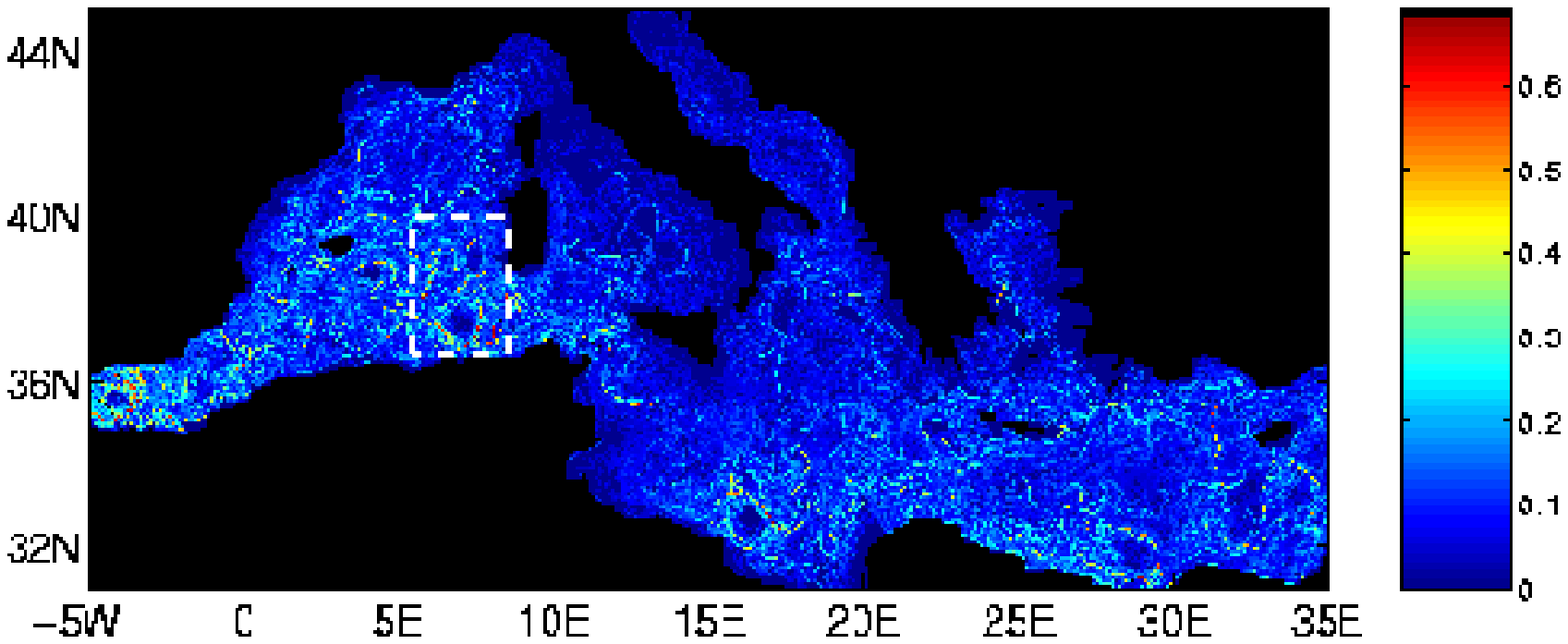}
\caption{FSLE spatial distribution for the whole Mediterranean on a specific
day (10th of June of the first simulation year). $\Delta x= \delta_0=
0.02$ degrees. Mesoscale structures and
vortices can be clearly seen. Units for FSLEs are day$^{-1}$. A zoom of the
indicated box is presented in fig.~\ref{fig:bigvortex}. 
}
\label{fig:fsleday161}
\end{figure}

\begin{figure}
\noindent\includegraphics[width=39pc]{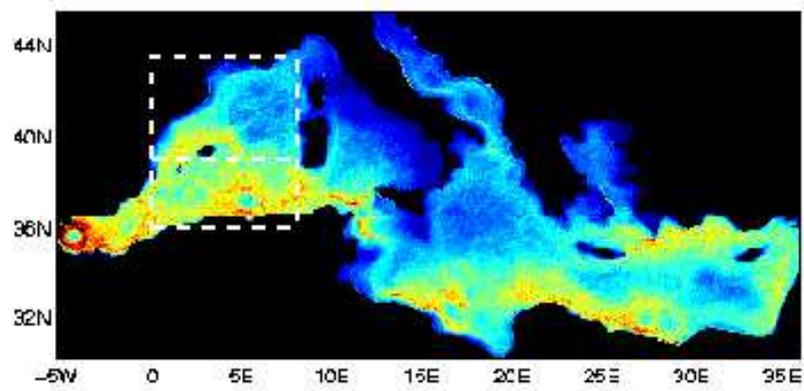}
\caption{Time average (for the first simulation year) of the FSLEs in the whole Mediterranean
basin. Geographical regions of different mixing activity appear. Colors as in
fig.~\ref{fig:fsleday161}.  }
\label{fig:fsleaverage}
\end{figure}

\begin{figure}
\includegraphics[width=20pc]{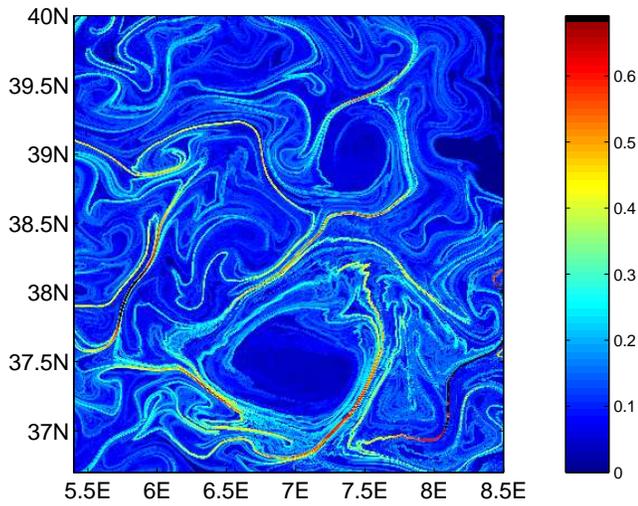}
\caption{Enlarged plot of the small box shown in fig.~\ref{fig:fsleday161}.
}
\label{fig:bigvortex}
\end{figure}

The spatial distribution of FSLEs for a particular day of the simulation (10th of
June of the first year of the data set) can be seen in Fig.~\ref{fig:fsleday161}.
Typical values are in the order of $0.1-0.6\ days^{-1}$, corresponding to mixing
times for mesoscale distances of $1.7-10\ days$. As observed in previous works,
maximum values of the distribution organize in lines \cite{Joseph2002,koh02} that
provide good approximations to repelling material lines (which are in turn stable
manifolds of hyperbolic moving points) \cite{Joseph2002}. These lines organize
the transport processes in the basin. Spatial structures ranging from the small
scales to the ones typical of mesoscale vortices are clearly identified.
Computing such picture for every day of the year and taking then the time
average, one can obtain a map of regions in the Mediterranean with different
mixing activity (fig.~\ref{fig:fsleaverage}). As expected, the Southern part of
the basin appears more active, especially close to the North African coast. In
fig.~\ref{fig:bigvortex} we show more in detail the FSLEs of the area in the
small box of fig.~\ref{fig:fsleday161}.  In the core of the eddies one has low
values of the FSLEs (i.e., low dispersion rates); on the contrary, the largest
values of the FSLEs can be found in the outer part of the eddies, where the
stretching of the fluid parcels is particularly important.

Note that in some regions of the vortex cores, chaotic tangles are still observed
as local maxima of the FSLE distribution. These maxima are, however, not strong.
In fact, even if the stretching is locally very high, the requirement for two
points to diverge for more than $\delta_f=110 km$ gives a low value of $\lambda$
to such finer structures, since such distance is bigger than the size of the
vortex that acts as a containing barrier most of the time. This is an example of
a useful property of the FSLE technique: it allows to restrict the analysis to
the structures relevant for transport among selected lengthscales only.

Additional sets of coherent structures and organizing lines can be obtained by
computing the FSLEs from trajectory integration {\sl backwards in time}. Maxima
in the new distribution identify lines of maximum compression, approximating
attracting material lines or unstable manifolds of hyperbolic moving points
\cite{Joseph2002}. Since stable and unstable manifolds cannot be crossed by
particle trajectories, such lines strongly constrain and determine fluid motion.

\begin{figure}
\includegraphics[width=20pc]{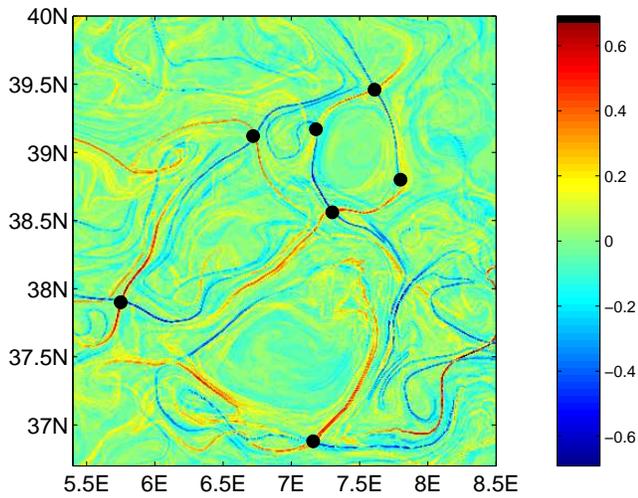}
\caption{FSLEs calculated from forward (displayed as positive values)
and backwards (displayed as negative values) integrations in time. A region with
strong mixing appears organized by a tangle of stretching and compressing
manifolds. Such lines organize the flow. The black dots indicate some of the
hyperbolic points that are located at the intersections of the lines. 
}
\label{fig:hyperbolic}
\end{figure}

Calculating in this way the FSLEs in a region of strong mixing, we unveil the
tangle of stretching and compressing lines in which vortices are embedded
(fig.~\ref{fig:hyperbolic}). These lines also define the directions of transport.
Lobes arising from intersections of stretching and compressing lines at a vortex
edge indicate where transport in and from the vortex takes place, whereas
tangencies among them provide barriers to transport. In Fig.~\ref{fig:hyperbolic}
some intersections of stretching and compressing lines are indicated as black
dots. These identify Lagrangian hyperbolic points (and their motion define
hyperbolic trajectories). Such points correspond to areas with strong mixing
activity: fluid is advected here along a compression line and then dispersed away
along the stretching line.

\begin{figure}
\includegraphics[width=20pc]{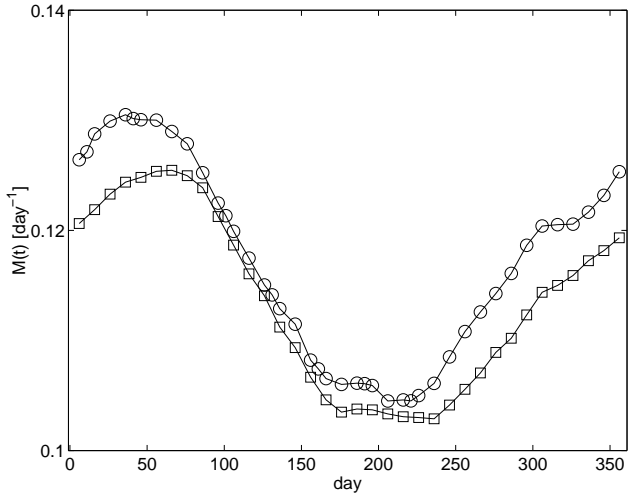}
\caption{Temporal evolution of the mixing measures $M_+(t)$ (circles), and of $M_\pm(t)$ (squares)
for the whole Mediterranean Sea during one simulation year. They display a very
similar behavior, with maximum values in winter.
}
\label{fig:seasonall}
\end{figure}

\begin{figure}
\noindent\includegraphics[width=20pc]{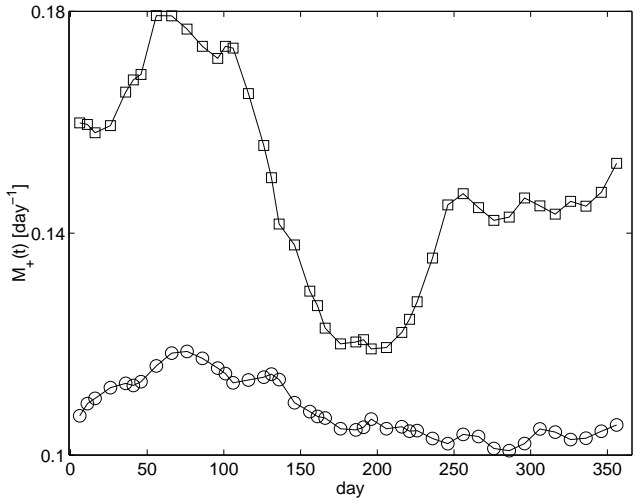}
\caption{$M_+(t)$ during one year for the Algerian current (squares), which
corresponds to the area delimited by the southern box in fig. \ref{fig:fsleaverage},
and
the north of the Balearic islands (circles), which is the northern box in
fig. \ref{fig:fsleaverage}.   
}
\label{fig:seasonloc}
\end{figure}

This dynamical picture suggests a quantitative measure of mixing in a prescribed
area $A$: one can define de quantity $M_\pm(t) \equiv
\left<\sqrt{\lambda_+\lambda_-} \right>_A$ where $\lambda_+$ and $\lambda_-$ are
the FSLEs in the forward and in the backwards time direction, and the average is
the spatial average over the area $A$. This quantity is large only where
hyperbolic points are present. The time dependence of this quantity when the area
$A$ is the whole Mediterranean is shown in fig.~\ref{fig:seasonall},
characterizing the seasonal variations of mixing. Maximum values,
of the order of $0.13$ $days^{-1}$, are attained in
winter. Because of the approximate incompressible character of the horizontal
flow, the temporal variations of forward and backward FSLEs are strongly
correlated, and one expects that the same information can be obtained from just
one of the FSLEs. Thus one can define a simpler measure of mixing in an area as
$M_+(t)=\left<\lambda_+\right>_A$. We show in fig.~\ref{fig:seasonall} that, as
expected, it contains essentially the same information as $M_{\pm}$, but at
variance with it, it could in principle be measured from floater experiments. It
is thus a more convenient characterization of mixing strength. As a further
example, we compare (fig. \ref{fig:seasonloc}) the temporal behavior of $M_+(t)$
in two regions (the areas North and South of the Balearic islands, corresponding
to the boxes in fig. \ref{fig:fsleaverage}) where
 we expect (from fig.~\ref{fig:fsleaverage}) to see a very different mixing activity.
The higher activity in the Southern part, where the Algerian current is present,
is confirmed. In addition,  seasonal fluctuations 
are smaller in the Northern part.

\section*{Conclusions} \label{sec:conclusions}

The FSLEs provide a direct method for computing simultaneously the mixing
activity and the coherent structures that control transport 
at a given scale. Interestingly, the method is based on the evolution of the
relative separation between two passive tracers, and thus numerical results
directly suggest floater-based experiments for verification. In this work we have
analyzed with this method horizontal velocity data from a Mediterranean computer
simulation. Different mixing behavior between geographical regions, and at
different seasons, is readily characterized. Finally, we point out that the
strong horizontal stirring associated with hyperbolic points should also have
important biological consequences. In the direction of recent works
\cite{martin03}, it would be interesting to compare the Lagrangian structures
presented here with productivity, patchiness, or other measures obtained from
biological distributions.

\section*{Acknowledgements}
We acknowledge financial support from MCyT of Spain and FEDER under projects
REN2001-0802-C02-01/MAR (IMAGEN) and BFM2000-1108 (CONOCE). C.L. is a {\it
Ram\'on y Cajal} research fellow (MCyT of Spain).

\end{document}